\documentclass[aps,superscriptaddress,nofootinbib,a4paper,11pt]{revtex4-2}
\usepackage{amsfonts,amsmath,amssymb}
\usepackage[compat=1.0.0]{tikz-feynman}
\usepackage{graphicx}
\usepackage{bm}
\usepackage[usenames,dvipsnames]{xcolor}
\usepackage{dcolumn}
\usepackage{comment}
\usepackage{enumerate}

\definecolor{venetianred}{rgb}{0.78, 0.03, 0.08}
\definecolor{midnightblue}{rgb}{0.1, 0.1, 0.44}
\definecolor{regalia}{rgb}{0.32, 0.18, 0.5}

\usepackage{hyperref} 
\hypersetup{
    colorlinks,
    linkcolor={venetianred},citecolor={midnightblue},urlcolor={regalia}
}

\def\gpqq{g_{\pi^0qq}}

\def\sm[#1]{{\scalebox{1.1}{$\scriptscriptstyle #1$}}}
\def\sv[#1]{{\scalebox{.9}{$\scriptscriptstyle #1$}}}
\def\sl[#1]{{\scalebox{1.25}{$\scriptscriptstyle #1$}}}
\def\per{\mkern-2mu \sl[\perp]} 
\def\perv{\mkern-2mu \sv[\perp]} 
\makeatletter
\newcommand{\npar}{\mathrel{\mkern+1mu \mathpalette\new@parallel\relax}}
\newcommand{\new@parallel}[2]{%
  \begingroup
  \sbox\z@{$#1 \scalebox{0.95}{T}$}
  \resizebox{!}{\ht\z@}{\raisebox{\depth}{$\m@th#1 ||$}}%
  \endgroup
}
\def\mf{{\mbox{\tiny MF}}}
\def\trD{\mathrm{Tr}_\sm[\rm D]}

\begin{document}

\title{\sc\Large{$\pi^0\rightarrow 2\gamma$ decay under strong magnetic fields in the NJL model}
\vspace*{5mm}}

\author{M\'aximo Coppola\,} 
\affiliation{CNEA, Departamento de F\'isica Te\'orica (DFTIFSC), 1429 Buenos Aires, Argentina}
\author{Daniel Gomez Dumm\,} 
\affiliation{IFLP, CONICET—Departamento de F\'isica, Facultad de Ciencias Exactas,
Universidad Nacional de La Plata, 1900 La Plata, Argentina}
\author{Norberto N. Scoccola\,} 
\affiliation{CNEA, Departamento de F\'isica Te\'orica (DFTIFSC), 1429 Buenos Aires, Argentina}
\affiliation{CONICET, 1033 Buenos Aires, Argentina \vspace*{1.5cm}}


\begin{abstract}
{\centering \large Abstract \par} %
We study the anomalous $\pi^0 \to \gamma\gamma$ decay under an external
uniform magnetic field in the framework of a two-flavor Nambu-Jona-Lasinio
model. It is seen that the  full decay width gets strongly reduced with the
external field, and that the differential width is almost independent of the
direction of the outgoing photons. We also find that the result for the
total width can be very well approximated by a simple expression obtained at
the lowest order in the chiral expansion, which is just a direct
extension to finite magnetic fields of the well-known $B=0$ result that
follows from the anomalous Wess-Zumino-Witten action. As a result, the
magnetic suppression of the width can be understood in terms of the
evolution of both the neutral pion mass and decay constant with the magnetic
field.
\end{abstract}

\maketitle
\clearpage


\section{Introduction}
\label{Sec:intro}

In the past few decades, there has been an increasing interest in the effect
of intense magnetic fields on the physics of strong
interactions~\cite{Kharzeev:2012ph,Andersen:2014xxa,Miransky:2015ava,Adhikari:2025104199}.
To a great extent, this is due to the impact that such large magnetic fields
might have in the study of the early
Universe~\cite{Vachaspati:1991nm,Grasso:2000wj}, in the analysis of high-energy 
noncentral heavy ion collisions~\cite{Skokov:2009qp,Voronyuk:2011jd},
and in the description of compact stellar objects like
magnetars~\cite{Duncan:1992hi,Kouveliotou:1998ze}. In addition,
magnetic fields have been shown to induce several interesting phenomena in
quantum chromodynamics (QCD), such as the chiral magnetic
effect~\cite{Kharzeev:2007jp,Fukushima:2008xe,Kharzeev:2015znc}, the
enhancement of the QCD quark-antiquark condensate (magnetic
catalysis)~\cite{Gusynin:1995nb}, the decrease of critical temperatures for
chiral restoration and deconfinement QCD transitions [inverse magnetic
catalysis (IMC)]~\cite{Bali:2011qj}, etc.

The study of strong interactions in the presence of external magnetic
fields is also useful to probe QCD dynamics, through, for instance, the
analysis of the effects of these fields on the properties of light hadrons.
The corresponding theoretical calculations usually represent a nontrivial
task, since they require to deal in general with QCD in a nonperturbative
regime. In view of this difficulty, magnetic effects have been studied
considering various effective theoretical approaches to strong interaction
dynamics. For example, the effect of intense external magnetic fields on
light pseudoscalar meson masses has been studied in the framework of
Nambu-Jona-Lasinio (NJL)-like
models~\cite{Fayazbakhsh:2012vr,Fayazbakhsh:2013cha,Liu:2014uwa,Avancini:2015ady,
Zhang:2016qrl,Avancini:2016fgq,Mao:2017wmq,GomezDumm:2017jij,Wang:2017vtn,Liu:2018zag,
Coppola:2018vkw,Mao:2018dqe,Avancini:2018svs,Coppola:2019uyr,Cao:2019res,Cao:2021rwx,Sheng:2021evj,
Avancini:2021pmi,Coppola:2023mmq}, quark-meson
models~\cite{Kamikado:2013pya,Ayala:2018zat,Ayala:2020dxs,Ayala:2023llp},
chiral perturbation theory
(ChPT)~\cite{Agasian:2001ym,Andersen:2012zc,Colucci:2013zoa}, path integral
Hamiltonians~\cite{Orlovsky:2013gha,Andreichikov:2016ayj}, effective chiral
confinement Lagrangians~\cite{Simonov:2015xta,Andreichikov:2018wrc} and QCD
sum rules~\cite{Dominguez:2018njv}. 
In addition, several results for the
spectrum of these mesons in the presence of background magnetic fields have
been obtained from lattice QCD (LQCD)
calculations~\cite{Bali:2011qj,Luschevskaya:2015bea,Luschevskaya:2014lga,Bali:2017ian,Ding:2020hxw,Ding:2022tqn,Endrodi:2024cqn}.

The effect of a magnetic field on the decay widths of light pseudoscalar
mesons is a related topic that has also been considered in the literature.
However, most of the existing works have been devoted to the analysis of the
leptonic decays of charged
pions~\cite{Nikishov:1964zza,Nikishov:1964zz,Bali:2018sey,Coppola:2018ygv,Coppola:2019idh,Coppola:2019wvh}.
The main conclusion of Refs.~\cite{Bali:2018sey,Coppola:2019wvh}, where
actual predictions for these processes are given, is that the presence of
the external field leads to a great enhancement of the corresponding decay
widths. On the other hand, the influence of a magnetic field on the
anomalous decay $\pi^0\to\gamma\gamma$ has received far less attention. 
In the past few years, this problem has been studied in
Refs.~\cite{Brauner:2017uiu,Adhikari:2024vhs} in the framework of ChPT,
where the validity of the analysis is restricted to values of $eB$ up to
$\sim m_\pi^2$. At this point, we should emphasize the importance of the
$\pi^0\to\gamma\gamma$ decay in the context of chiral theories. In fact, it
was through a careful analysis of the corresponding decay amplitude (in
vacuum) that anomalies were first
discovered~\cite{Adler:1969gk,Bell:1969ts}, and this led to the realization
that the U(1)$_A$ symmetry is not conserved in quantum field theories like
QCD. Having this in mind, it is clear that the understanding of how the
$\pi^0 \rightarrow \gamma \gamma$ decay width gets modified by the presence
of an external magnetic field appears to be a quite relevant issue.

The aim of the present work is to study the influence of a strong
magnetic field $\vec B$ on the $\pi^0$ anomalous decay using a
two-flavor NJL model. 
We recall that the NJL model is an effective
chiral quark model of QCD in which gluon degrees of freedom are
integrated out in favor of some local quark-antiquark interactions
that lead to the dynamical breaking of chiral symmetry. The model
is nonrenormalizable and, therefore, a regularization procedure
has to be adopted in order to deal with divergent quantities. The
choice of this procedure may be considered as part of the
definition of the model itself. 
In this work, we adopt a magnetic-field-independent regularization (MFIR)
method~\cite{Ebert:1999ht,Menezes:2008qt,Allen:2015paa}, which has
been shown to be convenient when dealing with magnetized
systems~\cite{Avancini:2019wed}. While in the NJL model quark
condensates emerge from the usual mean field approximation, mesons
are treated as excitations whose masses are obtained by
considering second-order corrections in the corresponding
bosonized action. The quark propagator in the presence of a
uniform magnetic field, considered in this work in the Schwinger
form~\cite{Andersen:2014xxa,Miransky:2015ava}, is used to evaluate
the $\pi^0$ polarization function and its decay amplitude to two
photons. As an additional ingredient, we consider the case of
$B$-dependent effective quark-quark coupling constants.
This possibility has been previously explored in effective
models~\cite{Ayala:2014iba,Farias:2014eca,Ferreira:2014kpa} in
order to reproduce the IMC effect observed
at finite temperature in LQCD calculations.

This paper is organized as follows. 
In Sec.~\ref{sec:theo}, we present the
theoretical formalism, introducing the magnetized two-flavor NJL-like
Lagrangian to be used in our calculations. Expanding the bosonized action in
powers of fluctuations around mean field values up to quadratic order, we
derive expressions for certain relevant quantities associated with the
$\pi^0\to \gamma\gamma$ decay width, such as quark masses, quark
propagators, the quark-pion coupling strength and the neutral pion mass.
In Sec.~\ref{sec:decay}, we calculate the magnetized $\pi^0\to \gamma\gamma$
decay width from the third-order expansion of the bosonized action. 
We also show that in the chiral limit, the resulting coupling constant
$g_{\pi^0\gamma\gamma}$ has a simple expression that turns out to be a
direct extension to finite magnetic field of the well-known $B=0$ result
that follows from the gauged Wess-Zumino-Witten action. 
In Sec.~\ref{sec:num}, we present and discuss our numerical results, while in
Sec.~\ref{sec:concl} we provide a summary of our work, together with our
main conclusions. We also include Appendices A, B and C to provide some
technical details of our calculations.


\vspace*{6mm}
\section{NJL model under strong magnetic fields: Mean field properties, $\pi^0$ mass and coupling to quarks}
\label{sec:theo}

The Lagrangian density of the NJL two-flavor model, including
couplings with an electromagnetic field, is given by
\begin{align}
{\cal L} \ = \ \bar{\psi}\left(i\,\rlap/\!D-m_0 \right)\psi
+ G \left[ \left(\bar{\psi} \psi\right)^2 + \left(\bar{\psi}\, i \gamma_5 \vec \tau \psi\right)^2 \right]
- \dfrac{1}{4} F_{\mu\nu} F^{\mu\nu} \ ,
\label{equno}
\end{align}
where $\psi=(u\ d)^{T}$, $\tau_i$ are Pauli matrices and $m_0$ is the current quark mass, which is assumed to be equal for $u$ and $d$ quarks.
The interaction between fermions and the electromagnetic field ${\cal A}_{\mu}$ is driven by the covariant derivative
\begin{align}
D_{\mu} \ = \ \partial_{\mu}+i\,\hat{Q}\mathcal{A}_{\mu}\ ,\label{covdev}
\end{align}
where $\hat{Q}=\mbox{diag}(q_u,q_d)$, with $q_u=2e/3$ and $q_d=-e/3$, 
while $e>0$ is the proton electric charge. 
In addition, Eq.~\eqref{equno} involves the
electromagnetic tensor $F^{\mu\nu} =
\partial^\mu {\cal A}^{\nu} - \partial^\nu {\cal A}^{\mu}$. The gauge field
$\mathcal{A}^{\mu} =A^\mu + a^{\mu}$ involves both the external
classical field $A^\mu$ and the photon degrees of freedom, given by
the dynamical quantum field $a^{\mu}$. We consider the particular case in which the external
electromagnetic field is given by an homogeneous stationary magnetic field
$\vec{B}$ orientated along the positive 3-axis.

Since we are interested in studying a meson decay, it is convenient to
bosonize the fermionic theory, introducing a scalar field $\sigma(x)$ and
pseudoscalar fields $\vec \pi(x)$ while integrating out fermion fields. The
bosonized action can be written as
\begin{align}
S_{\rm bos} \ = \ -i\ln\det\!\big(i\mathcal{D}\big)-\dfrac{1}{4G}\int d^{4}x\
\big[ \sigma(x)\,\sigma(x)+\vec{\pi}(x)\cdot\vec{\pi}(x) \big] + S_\sm[\rm EM] \ ,
\label{sbos}
\end{align}
with
\begin{align}
i\mathcal{D}_{x,y} \ = \ \Big[i\,\rlap/\!D_x-m_{0}- \sigma(x)-i\,\gamma_{5}\,
\vec \tau \cdot \vec \pi(x) \Big] \, \delta^{(4)}(x-y)\ ,
\label{dxx}
\end{align}
where a direct product to an identity matrix in color space is understood.
The pure electromagnetic field contribution to the effective action is given by
\begin{align}
S_\sm[\rm EM] \ = \ - \dfrac{1}{4} \int d^{4}x \ F_{\mu\nu} F^{\mu\nu} \ = \
-\dfrac{B^2}{2}\: V^{(4)} + \dfrac{1}{2} \int d^{4}x \ a_\mu(x) \Big(
\eta^{\mu\nu}\, \partial_\alpha \partial^\alpha - \partial^\mu \partial^\nu \Big)
a_\nu(x)\ ,
\end{align}
where the first and second terms on the rhs correspond to the classical and
propagating contributions, respectively. For the Minkowski metric, the
convention $\eta^{\mu\nu}=\mbox{diag}(1,-1,-1,-1)$ is adopted. Since we
consider a static and uniform external magnetic field, in what follows the
classical contribution to $S_\sm[\rm EM]$ will be dropped. Notice,
however, that this contribution could be relevant in some other contexts,
such as the study of charged pion superfluidity~\cite{Cao:2021gfk}.

We proceed by expanding the bosonized action in powers of fluctuations of
the bosonic fields around their mean field (MF) values. We assume that the
scalar field $\sigma(x)$ has a nontrivial translational invariant MF value
$\bar{\sigma}$, while vacuum expectation values of pion fields $\pi_i(x)$ are
$\bar{\pi}_i=0$ so as to respect the parity symmetry of vacuum. Thus, we
write
\begin{align}
\mathcal{D}_{x,y} \  =\ {\rm
diag}\big(\mathcal{D}_{x,y}^{\mf,\,u}\,,\,\mathcal{D}_{x,y}^{\mf,\,d}\big)\,
+\delta\mathcal{D}_{x,y}\ . \label{dxxp}
\end{align}
The MF piece, which is diagonal in flavor space, includes the interaction of
quark fields with the classical electromagnetic field $A^\mu$. One has
\begin{align}
\mathcal{D}_{x,y}^{\mf,\,f} \ = \ -i\,\left[ i\rlap/\partial_x-Q_{f}\, \rlap/\!A(x) -M \right]\, \delta^{(4)}(x-y) \ ,
\end{align}
where $M=m_0+\bar{\sigma}$ is the quark effective mass and $f=u,d$. On the
other hand, $\delta\mathcal{D}_{x,y}$ reads
\begin{equation}
\delta\mathcal{D}_{x,y} \ = \ i \left[ \hat Q \ \rlap/\!a(x) + \delta \sigma(x)+i\,\gamma_{5}\,
\vec \tau \cdot  \delta \vec \pi(x) \right] \, \delta^{(4)}(x-y) \ .
\label{dxx}
\end{equation}

Replacing in the bosonized effective action and expanding in powers of  meson fluctuations around the MF values, we get
\begin{align}
S_{\rm bos} \ = \ S_{\rm bos}^{\mf} + S_{\rm bos}^{\rm quad} + S_{\rm bos}^{\rm cub} + ...
\label{S_exp}
\end{align}
Here, the MF action per unit volume is given by
\begin{align}
\dfrac{S_{\mathrm{bos}}^{\mf}}{V^{(4)}}\ =\ - \dfrac{\bar{\sigma}^{2}}{4G}-\dfrac{iN_{c}}{V^{(4)}}
\sum_{f=u,d} \, {\rm Tr}\, \ln \big(i\mathcal{D}^{\mf,\,f}\big) \ ,
\label{seff}
\end{align}
where $N_c$ is the number of colors and the trace is understood to be
taken in both coordinate and Dirac spaces. Next, we introduce the MF quark
propagator in the presence of the classical magnetic field,
$\mathcal{S}^{\mf,\,f}=\big(i\mathcal{D}^{\mf,\,f}\big)^{-1}$. We choose
to use for this propagator the Schwinger form, given by
\begin{align}
\mathcal{S}_{x,y}^{\mf,\,f} \ = \ e^{i\Phi_{\!\sv[f]}(x,y)} \int_p \ e^{-i\,p(x-y)}
\, \bar{\cal S}_{f}(p)\ ,
\label{propx}
\end{align}
where
\begin{align}
\bar{\cal S}_{f}(p) \ = \ & -i\int_{0}^{\infty} d\rho\
\exp\left[ -i\rho\left( M^{2}-p_{\npar}^{2}+\vec{p}_{\per}^{\:2}\,\dfrac{\tan(\rho B_{f})}{\rho B_{f}}-i\epsilon \right) \right] \nonumber \\[2mm]
& \times \left[ \left( p_{\npar}\cdot\gamma_{\npar}+M \right)(1-s_f\,\gamma^{1}\gamma^{2}
\tan(\rho B_f))-\dfrac{\vec{p}_{\per}\cdot\vec{\gamma}_{\per}}{\cos^2(\rho B_f)} \right]\ ,
\label{sfp_schw}
\end{align}
with $B_f = |q_f B|$ and $s_f={\rm sign}(q_f B)$.
Here, we have defined ``parallel'' and ``perpendicular'' four-vectors
\begin{align}
p_{\npar}^{\mu} \ \equiv \ (p^0,0,0,p^3)\ , \qquad\quad
p_{\per}^{\mu} \ \equiv \ (0,p^1,p^2,0)  \ ,
\end{align}
and we have introduced a shorthand notation for the momentum integral
\begin{align}
\int_p \ \equiv \ \int \dfrac{d^{4}p}{(2\pi)^{4}}\ .
\end{align}
We have also used the notations
$p_{\npar}\cdot{\gamma}_{\npar} = p^0\gamma^0 - p^3\gamma^3$ and
$\vec{p}_{\per}\cdot\vec{\gamma}_{\per} = p^1\gamma^1 + p^2\gamma^2$.
The gauge dependence of the propagator is encoded in the so-called Schwinger
phase $\Phi_f(x,y)$ associated to the classical external field
\begin{align}
\Phi_f(x,y) \ = \ q_f \int_{x}^y d\xi_\mu \left[A^\mu(\xi) +  \dfrac{1}{2}\ F_\sm[A]^{\mu\nu}\, (\xi_\nu-y_\nu)\right] \ ,
\label{sp}
\end{align}
where $F_\sm[A]^{\mu\nu}$ is the corresponding electromagnetic strength tensor.

As usual, from the minimization condition $\partial S_{\mathrm{bos}}^{\mf}/\partial
\bar{\sigma}=0$, one obtains a gap equation, which can be written as
\begin{align}
M \ = \ m_0+4\, G M N_c \,I\ ,
\label{gapeqs}
\end{align}
with
\begin{align}
I \ = \ \frac{i}{2M}\, \sum_{f=u,d} \;\int_p \, \trD\, \bar{\cal S}_{f}(p)\ ,
\label{integral}
\end{align}
where $\trD$ stands for trace in Dirac space. Notice that Eq.~(\ref{gapeqs})
is a gauge-independent expression, as it should be. Performing the usual
deformation of the proper time integration paths (which effectively reduces
to the substitution $\rho\to -i\tau$)~\cite{Schwinger:1951nm}, one gets
\begin{align}
I \ = \ \frac{1}{8\pi^2}\, \sum_{f=u,d} \;\int_0^\infty \,
\frac{d\tau}{\tau^2}\; e^{-\tau M^2}\, \tau B_f\,\coth(\tau B_f)\ .
\end{align}
This is a divergent integral and has to be properly regularized. 
As stated in Sec.~\ref{Sec:intro}, we use the MFIR scheme: 
for a given unregularized quantity, the corresponding
(divergent) $B\to 0$ limit is subtracted and then added in a regularized
form. Thus, each quantity can be separated into a ``$B=0$'' term and a
``magnetic'' contribution. The regularized integral is given
by~\cite{Coppola:2019uyr}
\begin{align}
I^{\rm reg} \ = \ I^{0,{\rm reg}} \ + \ I^{\rm mag}\ ,
\end{align}
where
\begin{align}
I^{\rm mag} & \ = \ \frac{1}{8\pi^2}\, \sum_{f=u,d} \;\int_0^\infty \,
\frac{d\tau}{\tau^2}\; e^{-\tau M^2}\, \big[
\tau B_f\,\coth(\tau B_f) - 1\big]\nonumber \\
& \ = \ \frac{M^2}{8\pi^2}\, \sum_{f=u,d}\left[
\frac{\ln \Gamma(x_f)}{x_f}\, - \,\frac{\ln (2\pi)}{2x_f}\, + \, 1 \, -\,
\Big(1-\frac{1}{2x_f}\Big)\ln x_f\right]\ ,
\end{align}
with $x_f = M^2/(2B_f)$. In turn, the divergent ``$B=0$'' piece can be
regularized by introducing a 3D momentum cutoff $\Lambda$. One
has~\cite{Klevansky:1992qe}
\begin{align}
I^{0,{\rm reg}} \ = \ & \dfrac{1}{2 \pi^2} \left[
\Lambda \sqrt{ \Lambda^2 + M^2 } + M^2\;
\ln{\left( \dfrac{ M}{\Lambda + \sqrt{\Lambda^2 + M^2}}\right)}\right]\ .
\label{I0reg}
\end{align}
It is worth noticing that, in general, the ``$B=0$'' term still depends
implicitly on $B$ (e.g.\ through the values of the dressed quark masses
$M$), hence it should not be confused with the value of the studied
quantity at vanishing external field.

Next, we consider quadratic terms in the expansion of $S_{\rm bos}$ in
powers of $\delta\mathcal{D}_{x,y}$. We omit the description of charged
pion contributions, since charged pions cannot mix neither with the photon
nor the $\pi^0$. In contrast, neutral pions are in general expected to mix
with photons~\cite{Brauner:2017uiu}.
Here, we consider only lowest-order dynamical electromagnetic
contributions to the $\pi^0 \rightarrow \gamma\gamma$ decay.
Thus, we dismiss possible $\pi^0 - \gamma$ mixing effects. 
As usual, the $\pi^0$ effective propagator can be obtained from a resummation of quark loop
chains, using the random phase approximation. Since we are dealing with a
neutral particle, the contributions of Schwinger phases associated with
quark propagators [see Eq.~\eqref{propx}] cancel out, leading to gauge
invariant polarization functions that depend only on the difference $x-y$
(i.e., they are translationally invariant). If one performs a Fourier
transformation, the conservation of momentum implies that the polarization
function is diagonal in the momentum basis. Thus, in this basis we have
\begin{align}
S_{\pi^0} \ = \ - \dfrac{1}{2} \int_p \, \delta \pi^0(-q)
\left[ \dfrac{1}{2G} - \Pi_{\pi^0}(q_{\npar}^2,q_{\per}^2) \right] \delta \pi^0(q) \ ,
\end{align}
where
\begin{align}
\Pi_{\pi^0}(q_{\npar}^2,q_{\per}^2) \ = \ -i N_c \, \sum_f \, \trD
\left[ i \bar{\cal S}_{f}(p^{+}) \, i \gamma_5 \,
i \bar{\cal S}_{f}(p^-) \, i \gamma_5 \right] \ .
\label{pi0polfun}
\end{align}
Here $p^\pm = p \pm q/2$, while $\bar{\cal S}_f(p)$ is given in Eq.~\eqref{sfp_schw}.
Once again, employing the MFIR scheme we
regularize the above polarization function by separating $\Pi^{\rm
reg}_{\pi^0} = \Pi^{\rm mag}_{\pi^0} + \Pi^{\rm 0,reg}_{\pi^0}$. The
corresponding expressions are given in App.~\ref{app.pi0pol}. Choosing the
frame in which the $\pi^0$ meson is at rest, its mass can be obtained by
solving the equation
\begin{align}
\dfrac{1}{2G} - \Pi_{\pi^0}^{\rm reg}(m_{\pi^0}^2,0) \ = \ 0 \ .
\end{align}
As a final remark, we recall that the pion field wave function has to be
renormalized. Notice that the external magnetic field introduces a spatial
anisotropy reflected in the momentum dependence
$\Pi_{\pi^0}(q_{\npar}^2,q_{\per}^2)$, which distinguishes parallel from
perpendicular components. We introduce a (parallel) ``wave function
renormalization constant'' $Z_{\npar}^{1/2}$, defined by fixing the residue
of the two-point function at the pion pole. This constant, and the related
quark-pion coupling constant $\gpqq$, are given by~\cite{Coppola:2019uyr}
\begin{align}
Z_{\npar}^{-1} \ = \ \gpqq^{-2} \ = \ \dfrac{\partial \Pi_{\pi^0}}{\partial q_{\npar}^2}
\bigg\rvert_{\!\!{\tiny\begin{array}{l}
               q_{\npar}^2 = m_{\pi^0}^2 \\ \vec{q}_{\per}^{\:2} \! = 0
             \end{array}}} \ .
\label{zpar}
\end{align}

\section{$\pi^0\to \gamma\gamma$ decay width}
\label{sec:decay}

\subsection{Contributions to the decay amplitude}

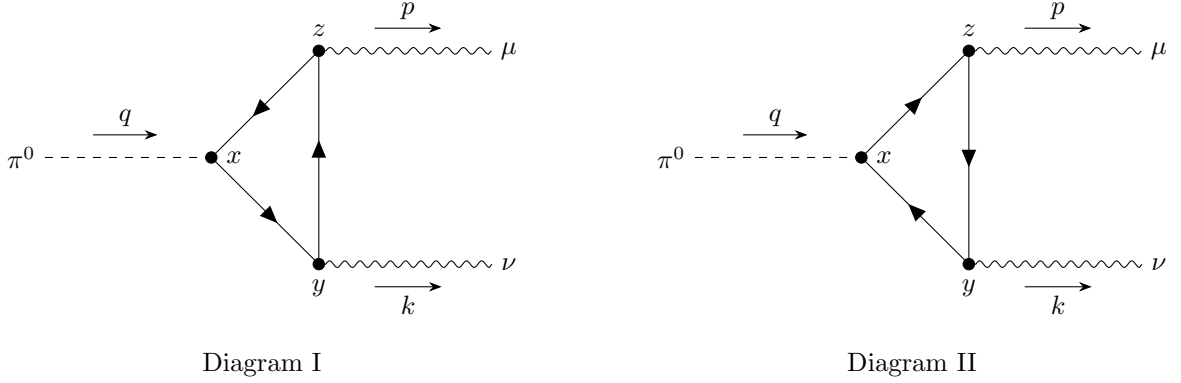
\begin{figure}[h]

\begin{minipage}{0.48\textwidth}\vspace{0pt}
\begin{center}
\begin{tikzpicture}
\begin{feynman}
\vertex (a) {\(\pi^0\)};
\vertex [right=2.5cm of a, dot, label=0:\(x\)] (b) {};
\vertex [above right=2cm of b, dot, label=above:\(z\)] (f1) {};
\vertex [below right=2cm of b, dot, label=below:\(y\)] (f2) {};
\vertex [right=2.5cm of f1] (c) {\(\mu\)};
\vertex [right=2.5cm of f2] (d) {\(\nu\)};
\diagram* {
(a) -- [scalar, momentum={[arrow shorten=0.3] \(q\)}] (b),
(b) -- [anti fermion] (f1) -- [anti fermion] (f2) -- [anti fermion] (b),
(f1) -- [photon, momentum={[arrow shorten=0.3] \(p\)}] (c),
(f2) -- [photon, momentum'={[arrow shorten=0.3] \(k\)}] (d),
};
\end{feynman}
\end{tikzpicture}
\\
Diagram I
\end{center}
\end{minipage}
\hfill%
\begin{minipage}{0.48\textwidth}\vspace{0pt}
\begin{center}
\begin{tikzpicture}
\begin{feynman}
\vertex (a) {\(\pi^0\)};
\vertex [right=2.5cm of a, dot, label=0:\(x\)] (b) {};
\vertex [above right=2cm of b, dot, label=above:\(z\)] (f1) {};
\vertex [below right=2cm of b, dot, label=below:\(y\)] (f2) {};
\vertex [right=2.5cm of f1] (c) {\(\mu\)};
\vertex [right=2.5cm of f2] (d) {\(\nu\)};
\diagram* {
(a) -- [scalar, momentum={[arrow shorten=0.3] \(q\)}] (b),
(b) -- [fermion] (f1) -- [fermion] (f2) -- [fermion] (b),
(f1) -- [photon, momentum={[arrow shorten=0.3] \(p\)}] (c),
(f2) -- [photon, momentum'={[arrow shorten=0.3] \(k\)}] (d),
};
\end{feynman}
\end{tikzpicture}
\\
Diagram II
\end{center}
\end{minipage}

\caption{One-loop Feynman diagrams that contribute to the
$\pi^0\rightarrow \gamma\gamma$ decay. Diagrams I and II
correspond to the amplitudes $i\mathcal{G}_f^{\mu \nu}(x,y,z)$ and
$i\mathcal{G}_f^{\nu \mu}(x,z,y)$, respectively. Internal
(external) arrows indicate the charge (momentum) flux. The
interaction points in configuration space are indicated with
labels $x$, $y$ and $z$.} \label{fig:diagrams}
\end{figure}

For a pion of momentum $q$ decaying into two photons of momenta $p$ and $k$,
the decay width is given by
\begin{align}
\Gamma_{\pi^0\gamma\gamma}(B) \ = \ \dfrac{1}{2 E_\pi} \dfrac{1}{2}
\int \dfrac{d^3p}{(2\pi)^3 2 p^0} \dfrac{d^3k}{(2\pi)^3 2 k^0}
\sum_{\lambda_p,\lambda_k=\pm} \dfrac{| i \mathcal{T} |^2}{V T} \ ,
\label{gamma0}
\end{align}
where $\lambda_p$ and $\lambda_k$ are the helicities of the outgoing photons,
while $V$ and $T$ are the volume and time interval in which the interaction
is active. 
At the end of the calculation, the limit $V, T \to \infty$ will be taken.
Note that a factor $1/2$ has been included to account for the
identical nature of the outgoing photons. The transition amplitude is
given by the relevant $\hat{\cal S}$-matrix element between the initial and
final states
\begin{align}
i \mathcal{T} \ = \ \langle \gamma(p,\lambda_p) \gamma(k,\lambda_k) | (\hat{\cal S}-1) | \pi^0(q) \rangle \ .
\end{align}
By properly expanding the fields in terms of creation and annihilation operators we get
\begin{align}
i \mathcal{T} \ = \ \epsilon_\mu(p,\lambda_p)^\ast \, \epsilon_\nu(k,\lambda_k)^\ast
\int d^4x \, d^4y \, d^4z \, e^{ipz} \, e^{iky} \, e^{-iqx} \:
\langle 0 | \, \textrm{T} [ J^\mu(z) J^\nu(y) J_5(x) ] \, | 0 \rangle  \ ,
\label{eqdos}
\end{align}
where $\epsilon_\mu(p,\lambda_p)$ and $\epsilon_\mu(k,\lambda_k)$ are the
photon polarization vectors, $\textrm{T}$ is the time-ordering operator and
the quark currents are given by
\begin{align}
J^\mu(x) \ = \ \bar{\psi}(x) (-\hat{Q} \gamma^\mu) \psi(x)\ , \qquad\qquad
J_5(x) \ = \ \bar{\psi}(x) (-\gpqq i \gamma_5 \tau_3) \psi(x) \ .
\end{align}
The connected three-point correlation function in Eq.~\eqref{eqdos} can be
obtained through Wick contractions. After taking trace over color, flavor
and coordinate spaces, we get
\begin{align}
i \mathcal{T}(q,p,k) \ = \ & \epsilon_\mu(p,\lambda_p)^\ast \, \epsilon_\nu(k,\lambda_k)^\ast \int d^4x \ d^4y \ d^4 z \
e^{i p z} \, e^{i k y} \, e^{-i q x} \nonumber \\
& \times\ \sum_{f=u,d} \left[ i \mathcal{G}_f^{\mu \nu}(x,y,z) + i \mathcal{G}_f^{\nu \mu}(x,z,y) \right] \ ,
\label{Tmom}
\end{align}
where
\begin{align}
i{\cal G}_f^{\mu \nu}(x,y,z) \ = \ i N_c \, \sum_{f=u,d} \, \trD \left[ (-\gpqq\, \kappa_f \,i \gamma_5) \, i \mathcal{S}_{x,z}^{\mf,\,f} \,
(-q_f \gamma^\mu) \, i \mathcal{S}_{z,y}^{\mf,\,f} \, (-q_f \gamma^\nu) \, i \mathcal{S}_{y,x}^{\mf,\,f} \right] \ .
\label{muno}
\end{align}
Here, the factors $\kappa_u=-\kappa_d=1$ arise from the Pauli matrix $\tau_3$ in the quark axial current $J_5(x)$.
The contributions $\mathcal{G}_f^{\mu \nu}(x,y,z)$ and $\mathcal{G}_f^{\nu \mu}(x,z,y)$ to the sum in Eq.~\eqref{Tmom}
correspond to the Feynman diagrams depicted in Fig.~\ref{fig:diagrams}.

Taking into account the form of the quark propagators given by
Eq.~\eqref{propx}, we obtain
\begin{align}
\mathcal{G}_f^{\mu \nu}(x,y,z) \ = \ & - \gpqq \, N_c \,
\kappa_f \, q_f^2 \; \exp\big[ i \left( \Phi_f(x,z) + \Phi_f(z,y) + \Phi_f(y,x) \right) \big]
\nonumber\\[2mm] & \times
 \int_{rst} e^{-i s (x-z)} e^{-i r (z-y)} e^{-i t (y-x)} \
 \trD  \left[i \gamma_5 \, i \bar {\cal S}_f(s) \, \gamma^\mu \,  i \bar {\cal S}_f(r)\,  \gamma^\nu \, i \bar {\cal S}_f(t) \right] \ .
\label{muno_v2}
\end{align}
It is easy to see that the sums of Schwinger phases appearing in the
amplitude are such that gauge-dependent pieces cancel out. Recalling that we
consider an homogeneous magnetic field $\vec{B}$ orientated along the
3-axis, we obtain the gauge-independent result
\begin{align}
\Phi_f(x,z) + \Phi_f(z,y) + \Phi_f(y,x) & \ = \
q_f (x_\mu-y_\mu)F^{\mu\nu}(z_\nu-x_\nu)/2 \nonumber \\
& \ = \ s_f\, B_f\, \epsilon_{ij3} (z_i-x_i)(x_j-y_j)/2\ .
\end{align}

At this point, we note that the two amplitudes in Eq.~\eqref{Tmom} 
are related by a charge conjugation transformation
$\hat{\cal C}$, namely
\begin{align}
\hat{\cal C} \; \mathcal{G}_f^{\nu \mu}(x,z,y) \; \hat{\cal C}^{-1} \ = \  \mathcal{G}_f^{\mu \nu}(x,y,z) \ .
\end{align}
Therefore, we can write
\begin{align}
i \mathcal{T}(q,p,k) \ = \ & -i \ \gpqq \, N_c \, \epsilon_\mu^\ast(p,\lambda_p) \, \epsilon_\nu^\ast(k,\lambda_k)
\nonumber \\[2mm]
&
\times\: \sum_{f=u,d} \kappa_f \, q_f^2 \sum_{\varepsilon=\pm 1}
\int_{rst} G^{(\varepsilon s_f)}(r,s,t) \ \trD \left[i \gamma^5 \, i \bar {\cal S}_{\varepsilon f}(t) \, \gamma^\nu\,
i \bar {\cal S}_{\varepsilon f}(r)\, \gamma^\mu \, i \bar {\cal S}_{\varepsilon f}(s) \right] \ ,
\label{tott}
\end{align}
where the subscript $\varepsilon f$ indicates that $s_f$ has to replaced by
$\varepsilon s_f$ in the corresponding propagator, as dictated by charge
conjugation. The function $G^{(\varepsilon s_f)}(s,r,t)$ is defined by
\begin{align}
G^{(\varepsilon s_f)}(r,s,t) \ = \ & \int d^4x \, d^4y \, d^4 z \: e^{i (p z + k y -q x)} \, e^{-is(x-z)-ir(z-y)-it(y-x)} \nonumber \\[2mm]
& \times \:  \exp\left[ i \, \dfrac{\varepsilon s_f B_f}{2} \, \epsilon_{ij3} (z_i-x_i)(x_j-y_j) \right] \ .
\label{ge}
\end{align}
After integration over space coordinates, one gets
\begin{align}
G^{(\varepsilon s_f)}(r,s,t) \ = \ & \dfrac{4}{B_f^2} \, (2\pi)^{10} \: \delta^{(4)}(q - p - k) \,
\delta^{(2)} (p_{\npar} + s_{\npar} - r_{\npar}) \, \delta^{(2)}(k_{\npar} + r_{\npar} - t_{\npar})  \nonumber\\[2mm]
& \hspace{-0.7cm} \times \exp \left\{ i \, \dfrac{2 \varepsilon s_f}{B_f}
\left[ (p^1+s^1-r^1)(k^2+r^2-t^2) - (k^1+r^1-t^1)(p^2+s^2-r^2) \right] \right\} \ .
\label{geexp}
\end{align}
As can be seen, while parallel momentum is conserved at each vertex,
the conservation of perpendicular momenta gets broken owing to Schwinger
phases, which mix the interaction points. Nevertheless, momentum is still
conserved for the overall $\pi^0\to \gamma\gamma$ process (as indicated by the
first Dirac delta), allowing us to define the decay amplitude $\mathcal{M}(q,p,k)$
as
\begin{align}
i \mathcal{T}(q,p,k) \ = \ (2\pi)^4 \, \delta^{(4)}(q-p-k) \, i \mathcal{M}(q,p,k) \ .
\label{calT}
\end{align}
It is convenient to write $\mathcal{M}(q,p,k)$ by indicating explicitly the contraction with
photon polarization vectors. We obtain
\begin{align}
i \mathcal{M}(q,p,k) \ = \  \epsilon_\mu(p,\lambda_p)^\ast\, \epsilon_\nu(k,\lambda_k)^\ast\:
\big(R^{\mu\nu}_u -R^{\mu\nu}_d\big) \ ,
\label{Mmunu}
\end{align}
where $R^{\mu\nu}_f$ is given by
\begin{align}
R_f^{\mu\nu} \ = \ & -i\, 16\, \pi^2  N_c \, \gpqq \,  \dfrac{q_f^2}{B_f^2} \,
\int_{rst} (2\pi)^{4}  \,
\delta^{(2)} (p_{\npar}+s_{\npar}-r_{\npar}) \: \delta^{(2)}(k_{\npar}+r_{\npar}-t_{\npar})
\nonumber\\[2mm]
& \times \sum_{\varepsilon=\pm 1} \ \exp \left\{ i\, \dfrac{2
\varepsilon s_f}{B_f}
\Big[ (p^1+s^1-r^1)(k^2+r^2-t^2) - (k^1+r^1-t^1)(p^2+s^2-r^2) \Big] \right\}  \nonumber \\[2mm]
& \times  \, \trD \left[i \gamma^5 \, i\bar{\cal
S}_{\varepsilon f}(s) \, \gamma^\mu \,  i\bar{\cal S}_{\varepsilon
f}(r)\, \gamma^\nu \, i\bar{\cal S}_{\varepsilon f}(t) \right] \ .
\label{rmunuf}
\end{align}

\subsection{Decay width calculation}

Taking into account Eq.~\eqref{gamma0} and the definitions in
Eqs.~\eqref{calT} and~\eqref{Mmunu}, the $\pi^0\to \gamma\gamma$ decay width is
given by
\begin{align}
\Gamma_{\pi^0\gamma\gamma}(B) \ = \ & \dfrac{1}{2 E_\pi} \, \dfrac{1}{2}
\int \dfrac{d^3p}{(2\pi)^3 2 p^0} \, \dfrac{d^3k}{(2\pi)^3 2 k^0} \
(2\pi)^4 \delta^{(4)}(q-p-k) \nonumber \\[2mm]
& \times \ \sum_{\lambda_p,\lambda_k=\pm} \big| \epsilon_\mu(p,\lambda_p)^\ast\,
\epsilon_\nu(k,\lambda_k)^\ast\,\big(R^{\mu\nu}_u - R^{\mu\nu}_d\big) \big|^2 \ .
\label{widthint}
\end{align}
This expression can be conveniently worked out in the pion rest frame, i.e.,
taking $q^\mu = (\, m_{\pi^0} ,\vec 0 \,)$. Momentum conservation leads then to
$\vec{p}=-\vec{k}$ and $p^0+k^0=m_{\pi^0}$. We assume that outgoing
photons can be taken at tree level in Quantum Electrodynamics (QED), neglecting higher-order
corrections that could arise from the presence of the external magnetic
field. Hence, we consider the usual on-shell conditions $p^2 = k^2 = 0$.

The calculation of $R^{\mu\nu}_f$ in the pion rest frame is outlined in
App.~\ref{app.decamp}. After some work, we arrive at the result
\begin{align}
R^{\mu\nu}_f \ = \  i \, \dfrac{\gpqq \, e^2 N_c\, M}{\pi^2} \
p^0 \,\epsilon^{0\mu\nu\lambda} \, p_\lambda \, \mathcal{I}_f \ ,
\label{Mmunu_final}
\end{align}
where we have used the convention $\epsilon^{0123}=+1$. The proper-time
integral $\mathcal{I}_f$ is given by
\begin{align}
\mathcal{I}_f \ = \ & \dfrac{q_f^2 \, B_f}{e^2} \,
\int_0^\infty d\tau_1 \int_0^\infty d\tau_2 \int_0^\infty d\tau_3 \:
\dfrac{1 +  t_1 t_3 - (t_1 + t_3) t_2}{(\tau_1+\tau_2+\tau_3)\,T_f}
\;\exp \big[-(\tau_1+\tau_2+\tau_3)\, M^2\big]
\nonumber\\[2mm] & \times \;
\exp\left[\dfrac{4 \tau_1 \tau_3\, (p^0)^2 + (\tau_1 +
\tau_3)\tau_2\,p_{\npar}^2}{\tau_1+\tau_2+\tau_3}\, - \,
\dfrac{(t_1 + t_3)t_2\, \vec p_{\per} \!\!\,^2}{T_f \, B_f}\right] \ ,
\label{Iftau}
\end{align}
where we have introduced the definitions $t_i = \tanh (\tau_i B_f)$ and
$T_f=t_1+t_2+t_3 + t_1 t_2 t_3$.

The contraction with photon polarization vectors can be carried out taking
into account the expression for $\epsilon_\mu(p,\lambda_p)$ in
App.~\ref{app.polB0}. One has
\begin{align}
\epsilon_\mu(p,\pm)^\ast\,
\epsilon_\nu(k,\pm)^\ast\,\epsilon^{0\mu\nu\lambda} \, p_\lambda \ = \ &
\mp\, ie^{\mp\, 2i\phi}\, p^0 \ ,\nonumber \\[2mm]
\epsilon_\mu(p,+)^\ast\,
\epsilon_\nu(k,-)^\ast\,\epsilon^{0\mu\nu\lambda} \, p_\lambda \ = \ &
\epsilon_\mu(p,-)^\ast\,
\epsilon_\nu(k,+)^\ast\,\epsilon^{0\mu\nu\lambda} \, p_\lambda \ = \
0 \ ,
\end{align}
where $\phi$ is an arbitrary phase. Therefore, from Eq.~\eqref{Mmunu_final}, we get
\begin{align}
\sum_{\lambda_p,\lambda_k=\pm} \big| \epsilon_\mu(p,\lambda_p)^\ast\,
\epsilon_\nu(k,\lambda_k)^\ast\,\big(R^{\mu\nu}_u - R^{\mu\nu}_d\big) \big|^2 \ = \
2\,m_{\pi^0}^4\,\alpha^2
\left(\dfrac{\gpqq \, N_c\, M}{\pi}\right)^2 \, (\mathcal{I}_u-\mathcal{I}_d)^2
\ ,
\end{align}
where $\alpha = e^2/(4\pi)$ is the fine structure constant.

As it is usually done, we can express the decay width in terms of an effective
pion-to-two-photon coupling $g_{\pi^0 \gamma\gamma}$, which will now depend
on the external field. Writing
\begin{align}
\Gamma_{\pi^0\gamma\gamma}(B) \ = \ \dfrac{\pi}{4} \, \alpha^2 \,
m_{\pi^0}^3 \, g_{\pi^0 \gamma\gamma}^2(B) \ ,
\label{GammaB}
\end{align}
and integrating over phase space in Eq.~\eqref{widthint}, we get
\begin{align}
g_{\pi^0 \gamma\gamma}^2(B) \ = \ \left( \dfrac{\gpqq \, N_c \, M}{2 \pi^2} \right)^2 \,
\int_0^1 d\cos\theta \ (\mathcal{I}_u-\mathcal{I}_d)^2 \ ,
\label{gpigg}
\end{align}
where $\theta$ is the angle between the magnetic field $\vec B$ and the
three-momentum $\vec{p}$ of one of the outgoing photons. One has then
$p_{\npar}^2=\vec{p}_{\per}^{\; 2}= m_{\pi^0}^2\sin^2\theta/4$. Finally, to
perform the integrals in Eq.~\eqref{Iftau} we find it convenient to introduce
new variables $x$, $y$ and $z$, defined by
\begin{gather}
\tau_1 \ = \ x z \ ,\qquad\qquad \tau_2 \ = \ y z\ , \qquad \qquad \tau_3 \ = \ (1-x-y) z \ .
\label{xyz}
\end{gather}
In this way, the proper-time integrals can be rewritten as
\begin{align}
\mathcal{I}_f \ = \ & \dfrac{q_f^2 \, B_f}{e^2} \,
\int_{0}^{\infty}dz \int_{0}^{1} dx \int_{0}^{1-x} dy \ \dfrac{z}{T_f}
\;\big[1 +  t_1 t_3 - (t_1 + t_3) t_2\big]
\nonumber\\[2mm] & \times \;
\exp\left\{-z\bigg[ M^2 -x(1-x-y)m_{\pi^0}^2\bigg]\, + \,
\dfrac{m_{\pi^0}^2}{4}\,\sin^2\theta\bigg[zy(1-y)-\dfrac{(t_1+t_3)t_2}{T_f\, B_f}\bigg] \right\} \ .
\label{Ifz}
\end{align}

As expected, the decay shows only an axially symmetric angular
distribution, since full rotational symmetry is broken by the external
magnetic field. The corresponding differential decay width reads
\begin{align}
\frac{d\Gamma_{\pi^0\gamma\gamma}(B)}{d\cos\theta} \ = \
\dfrac{\pi}{4} \, \alpha^2 \, m_{\pi^0}^3 \, \left( \dfrac{\gpqq \, N_c \, M}{2 \pi^2} \right)^2 \,
\big(\mathcal{I}_u-\mathcal{I}_d\big)^2 \ ,
\label{dgamma}
\end{align}
where the dependence with $\theta$ is given by Eq.~\eqref{Ifz}.

Our expression for the decay width can be evaluated in the limit of
vanishing external field, where  $t_i \rightarrow \tau_i B_f$ and
$T_f\to z B_f$. In this case, the $\theta$ dependence in Eq.~\eqref{Ifz} vanishes,
recovering spherical symmetry. We obtain
\begin{align}
g_{\pi^0 \gamma\gamma}(0) & \ = \ \gpqq \, \dfrac{N_c \, M}{6 \pi^2} \,
\int_{0}^{\infty}dz \int_{0}^{1} dx \int_{0}^{1-x} dy \ e^{-z \left[ M^2 - x(1-x-y) m_{\!   \pi^\sv[0]}^{2} \right] }
\nonumber\\
& \ = \ \gpqq \, \dfrac{N_c \, M}{3 \pi^2}
\left[\frac{1}{m_{\pi^0}}\,\arcsin\Big(\frac{m_{\pi^0}}{2M}\Big)\right]^2\ ,
\label{gpgg_0}
\end{align}
in agreement with the result obtained in Ref.~\cite{Klevansky:1994dk}.
Notice that in that reference the usual Feynman form of quark propagators
was considered.

\subsection{Chiral expansion}
\label{subsec:chiral}

As stated, it is also interesting to consider the anomalous $\pi^0$ decay
at the lowest order in the current quark mass expansion. Taking the
limit $m_{\pi^0} \rightarrow 0$ in Eq.~\eqref{Ifz} we have
\begin{align}
\mathcal{I}^{\,\rm ch}_f \ = \ \dfrac{q_f^2 \, B_f}{e^2}
\int_{0}^{\infty}dz \int_{0}^{1} dx \int_{0}^{1-x} dy \
\dfrac{z}{T_f} \; \left[ 1 +  t_1 t_3 - (t_1 + t_3) t_2 \right] \
e^{-z\,(M^{\rm ch})^2} \ ,
\end{align}
where $M^{\rm ch}$ is the ($B$-dependent) effective quark mass obtained in
the chiral limit $m_0=0$. This integral can be analytically performed,
leading to
\begin{align}
\mathcal{I}^{\,\rm ch}_f \ = \ \dfrac{q_f^2}{e^2}\,
\dfrac{1}{2\,(M^{\rm ch})^2} \ .
\end{align}
Replacing this expression into Eq.~\eqref{gpigg}, and noting that in
this case the integral over $\cos\theta$ can be trivially performed, we get
\begin{align}
g^{\rm ch}_{\pi^0 \gamma\gamma}(B) \ = \ \dfrac{N_c}{12\,\pi^2 f^{\,\rm ch}_{\pi^0}(B)} \ ,
\label{gch}
\end{align}
where we have made use of the generalized Goldberger-Treiman relation
$f^{\,\rm ch}_{\pi^0} \, g^{\rm ch}_{\pi^0 q q} = M^{\rm ch}$, which is
shown to be valid at finite $B$ (in the chiral limit) within the NJL
model~\cite{Coppola:2019uyr}. It is worth mentioning that, in fact, three
(vector and axial-vector) pion-to-vacuum form factors arise for the $\pi^0$
in the presence of the external magnetic field~\cite{Coppola:2018ygv}. 
Here, we denote by $f^{\,\rm ch}_{\pi^0}$ the one that reduces to the usual decay
constant $f_\pi$ at $B=0$. 
Moreover, Eq.~\eqref{gch} turns out to be a
direct extension of the well-known $B = 0$ result,
which follows from the anomalous gauged Wess-Zumino-Witten action introduced
in Refs.~\cite{Wess:1971yu,Witten:1983tw}, to a finite magnetic field.

In this way, the expression for the magnetized $\pi^0\to \gamma\gamma$ decay
width is found to be given by
\begin{align}
\Gamma_{\pi^0\gamma\gamma}^\sm[\rm LO](B) \ = \ \dfrac{\alpha^2}{64 \pi^3} \
\dfrac{m_{\pi^0}^3(B)}{[f^{\,\rm ch}_{\pi^0}(B)]^2} \ ,
\label{chlimit}
\end{align}
where the kinematical factor $m_{\pi^0}^3$ is taken to be nonzero, keeping
the lowest order in the chiral expansion. Clearly, for $B=0$ this equation
reduces to the well-known result obtained in ChPT~\cite{Itzykson:1980rh}.

\section{Numerical results}
\label{sec:num}

In this section we provide numerical results arising from our
calculation of the $\pi^0 \rightarrow \gamma\gamma$ decay width.
The values of related quantities, such as $m_{\pi^0}$ and $\gpqq$,
are estimated using the SU(2) version of the local NJL model. For
the model parameters we take the values $m_0 = 5.419$~MeV,
$\Lambda = 6.395$~MeV and $G\Lambda^2=2.136$, which (for vanishing
external magnetic field) correspond to a quark effective mass
$M=350$~MeV and a quark-antiquark condensate $\langle\bar f f
\rangle = (- 243.3\, \mbox{MeV})^3$. This parametrization, denoted
as S350, properly reproduces the empirical values of $m_{\pi^0}$
and $f_{\pi^0}$ in vacuum, namely $m_{\pi^0}=135$~MeV and $f_{\pi}
= 92.4$~MeV. To test the sensitivity of our results to the model
parametrization, we have also considered two alternative parameter
sets, denoted as S320 and S380, which also reproduce the
phenomenological values of $m_{\pi^0}$ and $f_\pi$ in vacuum, and
lead to quark effective masses $M = 320$ and 380 MeV,
respectively. The corresponding model parameters are listed in
Table~\ref{Table_Sets} together with the predictions for $\gpqq$
and $\Gamma_{\pi^0\gamma\gamma}$ at $B=0$. As one can see, the
values of $\Gamma_{\pi^0\gamma\gamma}$ are all compatible with the
experimental range $\Gamma^{\rm exp}_{\pi^0\gamma\gamma}=(7.72\pm
0.11)$~eV within less than 1\% accuracy.
\begin{table}[h!]
\centering
\begin{tabular}{c c c c c c c c}
\hline \hline
\qquad & \qquad $-\langle q \bar{q} \rangle^{1/3}$ \qquad & \qquad $m_0$ \qquad & \qquad $G \Lambda^2$ \qquad & \qquad $\Lambda$ \qquad & \qquad M \qquad & \qquad $\gpqq$ & \qquad $\Gamma_{\pi^0\gamma\gamma}$ \qquad \\
\qquad & \qquad MeV \qquad                                & \qquad MeV \qquad   & \qquad               \qquad & \qquad MeV \qquad & \qquad MeV \qquad & \qquad \qquad & \qquad eV \qquad \\
\hline
S320 \qquad & \qquad 246.9 \qquad & \qquad  5.185 \qquad & \qquad 2.138 \qquad & \qquad 639.5 \qquad & \qquad 320 \qquad & \qquad 3.396 & \qquad 7.67 \qquad \\
S350 \qquad & \qquad 243.3 \qquad & \qquad 5.418 \qquad & \qquad 2.252 \qquad & \qquad 613.4 \qquad & \qquad 350 \qquad & \qquad 3.720 & \qquad  7.65 \qquad \\
S380 \qquad & \qquad 241.4 \qquad & \qquad 5.543 \qquad & \qquad 2.366 \qquad & \qquad 596.1  \qquad & \qquad 380 \qquad & \qquad  4.046 & \qquad 7.65 \qquad \\
\hline \hline
\end{tabular}
\caption{Model parameters and predictions for $\gpqq$ and
$\Gamma_{\pi^0\gamma\gamma}$ at $B=0$.}
\label{Table_Sets}
\end{table}

In addition, since local NJL models fail to describe the IMC effect observed
in LQCD at finite temperature, we also consider a NJL Lagrangian with a
magnetic-field-dependent four-quark interaction coupling $G(B)$, so as 
to account for the effect of the magnetic field on sea quarks. For
definiteness, we adopt for $G(B)$ the form proposed in
Ref.~\cite{Avancini:2016fgq}, namely
\begin{equation}
G(B) \ =\ G \,\mathcal{F}(B) \ ,
\label{gdeb}
\end{equation}
where
\begin{equation}
\mathcal{F}(B) \ =\ \kappa_1 + (1-\kappa_1)\,e^{-\kappa_2(eB)^2} \ ,
\label{FeB}
\end{equation}
with $\kappa_1= 0.321$ and $\kappa_2= 1.31$~GeV$^{-2}$. Even though $G(B)$
does not appear explicitly in our expressions for the decay width
$\Gamma_{\pi^0\gamma\gamma}$, the $B$ dependence affects the values of $M$,
$m_{\pi^0}$ and $\gpqq$, which implicitly depend on the four-quark coupling
strength.

\begin{figure}[htb]
\centering{}\includegraphics[width=0.95\textwidth]{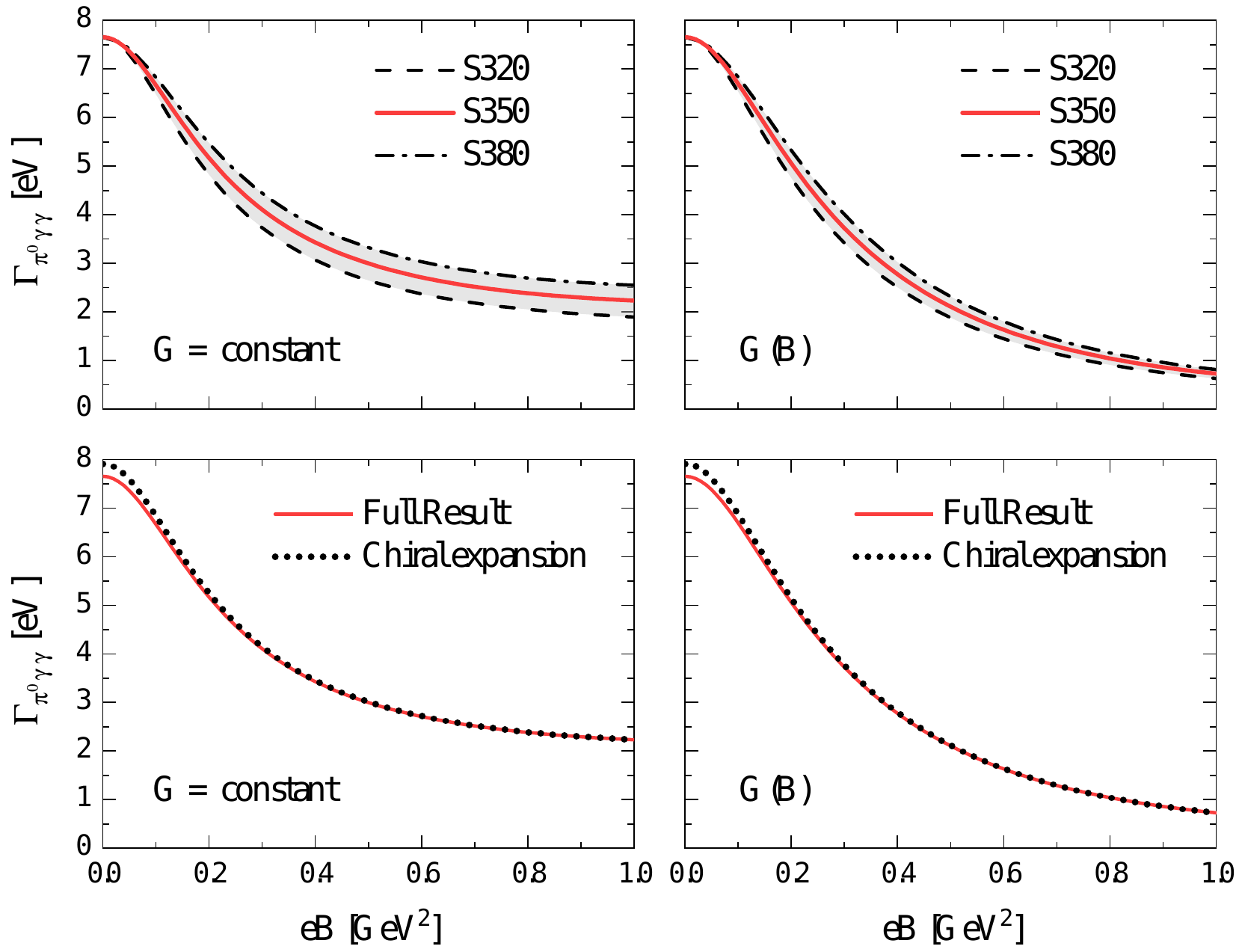}
\caption{$\pi^0 \rightarrow \gamma \gamma$ decay width as a function
of $eB$, for $G = {\rm constant}$ (left) and $G(B)$ (right). Solid red lines
correspond to parameter set S350. Band limits in upper panels correspond to
S320 (dashed lines) and S380 (dash-dotted lines). Dotted lines in lower panels
correspond to the lowest-order chiral expansion in Eq.~\eqref{chlimit}.}
\label{fig:Gamma_NM}
\end{figure}

Our numerical results for the $\pi^0 \rightarrow \gamma\gamma$ decay
width in the presence of the external magnetic field are collected in
Fig.~\ref{fig:Gamma_NM}. In the upper panels, we show our results for
$\Gamma_{\pi^0\gamma\gamma}$ for the three parameter sets of
Table~\ref{Table_Sets}, considering both a constant coupling $G$ (left) and
a magnetic-field-dependent coupling $G(B)$ (right). As seen, the NJL model
predicts a rather strong suppression of the width, which is even more
striking in the case of $G(B)$. We can also observe that this feature is
basically independent of the model parametrization. Before discussing the
possible origin of this suppression, it is convenient to compare the above
results with the predictions arising from Eq.~(\ref{chlimit}), obtained
at the lowest order in the current quark mass expansion. This
comparison is shown in the lower panels of Fig.~\ref{fig:Gamma_NM}
for the parameter set S350. As seen from the graphs, our calculations show
that for both $G={\rm constant}$ (left) and $G(B)$ (right), the
result in Eq.~\eqref{chlimit} is actually a good approximation to the full
width. Indeed, the accuracy of this approximation is found to be less than
4\% at $B=0$, becoming even smaller as the magnetic field increases.

\begin{figure}[h]
\centering{}\includegraphics[width=0.8\textwidth]{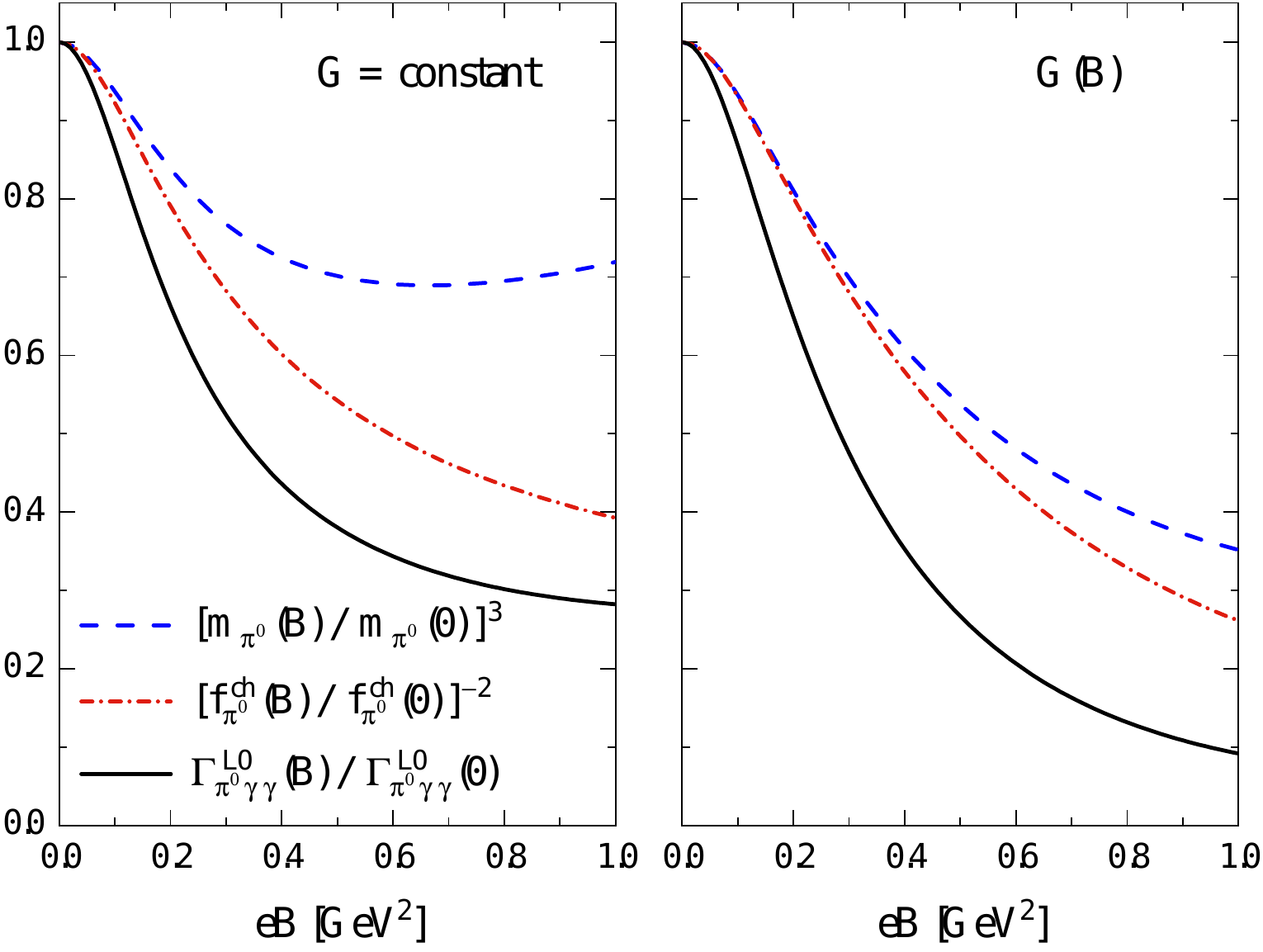}
\caption{Cubed $\pi^0$ mass (dashed lines), inverse squared pion decay
constant (dashed-dotted lines) and $\pi^0\to\gamma\gamma$ decay width (solid
lines) at lowest order in ChPT as functions of $eB$.
All three quantities are normalized with respect to their values at $B=0$.
The cases of a constant coupling $G$ and a $B$-dependent coupling $G(B)$ are
shown in left and right panels, respectively. All values correspond to the
parameter set S350.}
\label{fig:mfpiG}
\end{figure}

The similarity between the numerical results for
$\Gamma_{\pi^0\gamma\gamma}(B)$ and $\Gamma_{\pi^0\gamma\gamma}^\sm[\rm LO](B)$
can be used as a hint to trace the origin of the width suppression with the
external magnetic field. In this sense, we can take into account the fact
that at the lowest order in the chiral expansion, there is a simple
relation between the $\pi^0\to\gamma\gamma$ width, the pion mass and the
pion decay constant. The behavior of these three quantities as functions of
$B$ is shown in Fig.~\ref{fig:mfpiG} for set S350. Looking at these curves,
and considering Eq.~\eqref{chlimit}, it becomes clear that the width
suppression originates from the joint effect of the decrease of the neutral
pion mass and the increase of $f_{\pi^0}^{\rm ch}$ with the magnetic field.
In fact, it is seen that both effects turn out to be numerically important.
We note in this sense that in Ref.~\cite{Adhikari:2024vhs}, where the width
calculation was performed within the framework of ChPT, only the magnetic 
behavior of the pion mass was taken into account.
It is interesting to observe that the magnetic behavior of the pion decay
constant can be qualitatively understood taking into account the generalized
Gell-Mann-Oakes-Renner relation~\cite{Coppola:2019uyr} $(m_{\pi^0} f^{\,\rm
ch}_{\pi^0})^2 = -m_0 \langle \bar{u}u+\bar{d}d \rangle^{\rm ch} /2$ (valid
at lowest order in the current quark mass $m_0$), which explains the
enhancement of $f^{\,\rm ch}_{\pi^0}$ with $B$ in terms of both the magnetic
catalysis effect and the decrease of $m_{\pi^0}$ as $B$ increases. Regarding
the use of $G(B)$, it is seen from Fig.~\ref{fig:mfpiG} that magnetic
effects on $m_{\pi^0}$ and $f_{\pi^0}^{\,\rm ch}$ are more pronounced when
considering $G(B)$ in comparison with the case of a constant $G$. This leads
to a stronger suppression of the width for $G(B)$ that also holds away from
the chiral limit, as shown in Fig.~\ref{fig:Gamma_NM}.

Finally, the angular dependence of the decaying photon rate can be analyzed
from Eq.~(\ref{dgamma}). Though in principle the differential width
$d\Gamma(B)/d\cos\theta$ depends on the polar angle $\theta$ between the
magnetic field and one of the photon three-momenta, it is seen that the
$\theta$-dependent contribution is suppressed by powers of the ratio
$(m_{\pi^0}/M)^2$. Indeed, from our numerical calculations we find that for
values of $eB$ up to 1~GeV$^2$, the differential width varies less than
0.1\% within the full range of values of $\theta$.

\section{Summary and conclusions}
\label{sec:concl}

In this work we have studied the anomalous $\pi^0 \to \gamma\gamma$ decay in
the presence of a static uniform external magnetic field. Our analysis has
been performed in the framework of a two-flavor Nambu-Jona-Lasinio effective
model for low-energy QCD dynamics, in which pions are treated as quantum
fluctuations in the random phase approximation. The ultraviolet divergences
associated with the nonrenormalizability of the model have been regularized
using the magnetic-field-independent regularization method, which has been
shown to be free from unphysical oscillations and to reduce the dependence
of the results on the model parameters~\cite{Avancini:2019wed}.
Additionally, we have explored the possibility of using a 
magnetic-field-dependent coupling constant $G(B)$ to account for the effect 
of the magnetic field on sea quarks.

Our calculations indicate that the $\pi^0\to \gamma\gamma$ decay width gets
strongly reduced with the external field, a fact that can be more easily
understood by considering the current quark mass expansion. At
leading order in $m_{\pi^0}$, we find that the result for the full width
reduces to a simple expression, which turns out to be just a direct extension 
of the well-known $B = 0$ result that follows from the anomalous gauged 
Wess-Zumino-Witten action to a finite magnetic field. 
That expression linearly relates
$\Gamma_{\pi^0\gamma\gamma}^\sm[\rm LO]$ with the cube of the pion mass and the
inverse of the squared pion decay constant, $(f_{\pi^0}^{\,\rm ch})^{-2}$,
see Eq.~\eqref{chlimit}. Since $m_{\pi^0}$ decreases while $f_{\pi^0}^{\,\rm
ch}$ increases with $B$, the joint effect implies a strong suppression of
$\Gamma_{\pi^0\gamma\gamma}^\sm[\rm LO]$ with the external field. Our analysis
also shows that the effects of both the behavior of $m_{\pi^0}$ and
$f_{\pi^0}^{\,\rm ch}$ with $B$ contribute significantly to the width
reduction. Moreover, it is seen that this qualitative analysis can be
extrapolated away from the chiral limit: from our numerical results, it is
found that the calculation of $\Gamma_{\pi^0\gamma\gamma}$ at the
lowest order in the chiral expansion is actually a good approximation to
the full calculation, with the differences lying below 4\% for values of $eB$ up
to 1~GeV$^2$. The full analysis is found to be basically independent of the
model parametrization, within phenomenologically acceptable parameter
ranges. Regarding the use of a $B$-dependent four-quark coupling, we observe
that magnetic effects on $m_{\pi^0}$ and $f_{\pi^0}$ are more pronounced
when considering $G(B)$ with respect to the case of constant $G$. This leads
to an even stronger suppression of the $\pi^0\to\gamma\gamma$ width.

As for the angular dependence of the decaying photons, an anisotropic
angular distribution could be, in principle, expected, since rotational
symmetry is broken by the magnetic field. Nevertheless, we find that
angular-dependent contributions to the differential decay width
$d\Gamma(B)/d\cos\theta$ get suppressed by powers of the ratio
$(m_{\pi^0}/M)^2$, leading to nearly perfect isotropy: for
magnetic fields as large as $eB=1$~GeV$^2$, $d\Gamma(B)/d\cos\theta$ varies
less than 0.1\% within the full range of values of $\theta$.

In this work, we have considered only leading-order dynamical electromagnetic
contributions, taking vacuum dispersion relations for the outgoing photons
and dismissing possible effects arising from $\pi^0 - \gamma$ mixing.
We leave the analysis of the effect of the corresponding corrections for
future studies.


\begin{acknowledgments}
This work has been partially funded by CONICET (Argentina) under Grant
No.~PIP 2022--2024 GI-11220210100150CO, by ANPCyT (Argentina) under Grants
No.~PICT20-01847 and No.~PICT22-03-00799, and by UNLP (Argentina) under
Project No.~X960.
\end{acknowledgments}

\vspace*{5mm}
\appendix

\section{$\pi^0$ polarization function}
\label{app.pi0pol} 

The regularized form of the $\pi^0$ polarization function,
Eq.~\eqref{pi0polfun}, in the MFIR scheme can be written as
\begin{align}
\Pi^{\rm
reg}_{\pi^0}(q_{\npar}^2,q_{\per}^2) \ = \
\Pi^{\rm 0,reg}_{\pi^0}(q^2) + \Pi^{\rm mag}_{\pi^0}(q_{\npar}^2,q_{\per}^2) \ .
\end{align}
Here, the regularized ``$B=0$'' contribution, obtained using a 3D
momentum cutoff $\Lambda$, can be written as
\begin{align}
\Pi^{\rm 0,reg}_{\pi^0}(q^2) \ = \ 2N_c \left[ I^{0,{\rm reg}} - q^2 I^{0,{\rm reg}}_{2}(q^2) \right] \ ,
\end{align}
where $I^{0,{\rm reg}}$ is given by Eq.~(\ref{I0reg}), while $I_{2}^{0,{\rm
reg}}$ reads~\cite{Klevansky:1992qe,Coppola:2019uyr}
\begin{align}
I_2^{\rm 0,reg}(q^2) \ = \ & \dfrac{1}{4 \pi^2} \int_0^1 dy \left[
\dfrac{\Lambda}{\sqrt{\Lambda^2+ M^2 - y(1-y) q^2}}\, +\, \ln{ \dfrac{
{\sqrt{M^2 - y(1-y) q^2}}}{\Lambda + \sqrt{\Lambda^2 +
M^2 - y(1-y) q^2}}} \right] .
\end{align}
On the other hand, the magnetic contribution can be written as~\cite{Coppola:2019uyr}
\begin{align}
\Pi^{\rm mag}_{\pi^0}(q_{\npar}^2,q_{\per}^2) \ = \ & \dfrac{N_c}{4\pi^2} \sum_{f=u,d} \int_0^\infty dz\,\int_0^1 dy \
e^{ -z \left[ M^2 - y(1-y) q_{\sv[\parallel]}^2 \right] } \nonumber\\[2mm]
& \times \bigg\{ e^{-\omega_f(y,z)\,\vec{q}_{\perv}^{\:2}} \,
\bigg[ \left( M^2+\dfrac{1}{z}+y(1-y) q_{\npar}^2 \right) \,\dfrac{B_f}{\tanh(z B_f)} \nonumber\\[2mm]
& + \ \dfrac{B_f^2}{\sinh^2(z B_f)} \left(1-\omega_f(y,z)\,\vec{q}_{\per}^{\:2}\right) \bigg] 
- \dfrac{ e^{-zy(1-y)\,\vec{q}_{\perv}^{\:2}} }{z}
\left[ M^2+\dfrac{2}{z}+y(1-y) q^2 \right] \bigg\} \ ,
\end{align}
where
\begin{align}
\omega_f(y,z) \ = \ \dfrac{\sinh(yzB_f)\,\sinh[(1-y)zB_f]}{B_f\,\sinh(zB_f)} \ .
\end{align}
For a pion at rest, this expression can be rewritten in terms of the
digamma function, see Ref.~\cite{Coppola:2023mmq} for details.

\section{Computation of the $\pi^0\to \gamma\gamma$ decay amplitude}
\label{app.decamp} 

To evaluate the decay amplitude we need to calculate the tensor
$R^{\mu\nu}_f$ given by Eq.~\eqref{rmunuf}. Taking into account the
Schwinger form of quark propagators in Eqs.~\eqref{propx} and~\eqref{sfp_schw},
after integration over the ``parallel'' coordinates $s_{\npar}$ and
$t_{\npar}$ we obtain
\begin{align}
R_f^{\mu\nu} \ = \ & i\,16 \pi^2 N_c \, \gpqq \,  \dfrac{q_f^2}{B_f^2} \,
\int_0^\infty d\tau_1 \int_0^\infty d\tau_2 \int_0^\infty d\tau_3 \: e^{-(\tau_1+\tau_2+\tau_3)M^2}
\nonumber\\[2mm]
& \times \,
\int \dfrac{d^2r_{\npar}}{(2\pi)^2}\
e^{\tau_ 1 (r_{\sv[\parallel]}- p_{\sv[\parallel]})^2 + \tau_2\, r_{\sv[\parallel]}^2 + \tau_3 (r_{\sv[\parallel]}+ k_{\sv[\parallel]})^2}
\sum_{\varepsilon=\pm 1} \: {\cal N}^{\mu\nu}_{\varepsilon f} \ ,
\label{Mmunu2}
\end{align}
where
\begin{align}
 {\cal N}^{\mu\nu}_{\varepsilon f} \ = \ &
\int \dfrac{d^2r_{\per}}{(2\pi)^2}\, \dfrac{d^2s_{\per}}{(2\pi)^2} \, \dfrac{d^2t_{\per}}{(2\pi)^2} \
\exp \left\{- \dfrac{1}{B_f}
\left( t_1 \, {\vec s}_{\per}^{\;2} + \, t_2\, {\vec r}_{\per}^{\;2} + \, t_3 \,{\vec t}_{\per}^{\;\,2}
\right) \right\}
\nonumber\\[2mm]
& \times \, \exp\left\{ i \, \dfrac{ 2 \varepsilon s_f}{B_f} \left[ ( p^1+s^1-r^1)(k^2+r^2-t^2) -
(k^1+r^1-t^1)(p^2+s^2-r^2) \right] \right\}
\, T^{\mu\nu}_{\varepsilon f}\ ,
\label{nmunu}
\end{align}
with $t_i = \tanh(\tau_i B_f)$. Here, $T^{\mu\nu}_{\varepsilon f}$
stands for the Dirac trace
\begin{equation}
T^{\mu\nu}_{\varepsilon f}\ = \
\trD \left[i \gamma^5 \, i C_{\varepsilon f}(r_{\npar}-p_{\npar},s_{\per},t_1) \, \gamma^\mu \, i
C_{\varepsilon f}(r_{\npar},r_{\per},t_2) \, \gamma^\nu \, i
C_{\varepsilon f}(r_{\npar}+k_{\npar},t_{\per},t_3) \right] \ ,
\label{tmunu}
\end{equation}
where
\begin{align}
C_{\varepsilon f}(u_{\npar},v_{\per},t_i) \ = \ &
(u_{\npar}\cdot\gamma_{\npar} + M)\,\big[1 + i\varepsilon
s_f\,\gamma^1\gamma^2 t_i\big] \,- \,
\vec v_{\per}\cdot\vec\gamma\,(1-t_i^2)\ .
\end{align}

From explicit computation of the traces in Eq.~\eqref{tmunu} we obtain
\begin{align}
T^{00} \ = \ & -8\,\varepsilon s_f M T_f\, p^0 p^3 \ ,
\nonumber \\[2mm]
T^{11} \ = \ & T^{22} \ = \ 8\,\varepsilon s_f M \,[t_1+t_3-t_2(1+t_1 t_3)]\, p^0 p^3 \ ,
\nonumber \\[2mm]
T^{33} \ = \ & -8\,\varepsilon s_f M T_f\, p^0 (2 r^3 -p^3)\ ,
\nonumber \\[2mm]
T^{01} \pm i T^{02} \ = \ & \pm 4\, M f_1^\pm f_2^\pm f_3^\mp
\,\big[f_1^\mp (s^1\pm is^2) - f_3^\pm (t^1\pm it^2)\big]\,p^3\ ,
\nonumber \\[2mm]
T^{10} \pm i T^{20} \ = \ & \mp 4\, M f_1^\pm f_2^\mp f_3^\mp
\,\big[f_1^\mp (s^1\pm is^2) - f_3^\pm (t^1\pm it^2)\big]\,p^3\ ,
\nonumber \\[2mm]
T^{03} + T^{30} \ = \ & -16\, \varepsilon s_f M T_f\, r^0 p^0\ ,
\nonumber \\[2mm]
T^{03} - T^{30} \ = \ & 8 M \varepsilon s_f \Big\{  T_f\, [M^2 + p_{\npar}^2 - r_{\npar}^2
+ 2 p^3(p^3+r^3)]\, + t_1 (1-t_2^2)(1-t_3^2) (\vec r_{\per} \cdot \vec t_{\per})
\nonumber \\[2mm]
& + t_2 (1-t_1^2)(1-t_3^2) (\vec s_{\per} \cdot \vec t_{\per})
+ t_3 (1-t_1^2)(1-t_2^2) (\vec r_{\per} \cdot \vec s_{\per}) \Big\}
\nonumber \\[2mm]
& + 8\, i M \epsilon^{ij3} \Big\{ r^i t^j (1-t_2^2)(1-t_3^2) +
t^i s^j (1-t_1^2)(1-t_3^2) + s^i r^j (1-t_1^2)(1-t_2^2)\Big\}\ ,
\nonumber \\[2mm]
T^{12} \ = \ & - T^{21} \ = \ 8\,iM \,[1+t_1 t_3-t_2(t_1+t_3)]\, p^0 p^3 \ ,
\nonumber \\[2mm]
T^{13} \pm i T^{23} \ = \ & \pm 4\, M f_1^\pm f_2^\mp \Big\{
f_3^\mp \,\big[f_1^\mp (s^1\pm is^2) + f_3^\pm (t^1\pm it^2)\big]
-2 f_2^\pm f_3^\pm (r^1\pm ir^2)\Big\}\,p^0\ ,
\nonumber \\[2mm]
T^{31} \pm i T^{32} \ = \ & \mp 4\, M f_2^\pm f_3^\mp \Big\{
f_1^\pm \,\big[f_1^\mp (s^1\pm is^2) + f_3^\pm (t^1\pm it^2)\big]
-2 f_2^\mp f_1^\mp (r^1\pm ir^2)\Big\}\,p^0\ ,
\end{align}
where we have defined $f_i^\pm = 1 \pm \varepsilon s_f\, t_i\,$ and
$T_f=t_1+t_2+t_3+t_1t_2t_3$. We have also considered the usual photon on-shell
conditions $k^2=p^2=0$ and taken into account the fact that in the
rest frame of the pion one has $k^0 = p^0$, $k^3=-p^3$.

To calculate the integrals over perpendicular momenta in Eq.~\eqref{nmunu},
notice that the traces include up to quadratic powers of $r_{\per}$,
$s_{\per}$ and $t_{\per}$ coordinates. In fact, we are left with Gaussian
integrals that can be straightforwardly performed. Let us define
\begin{align}
I_{\per} (u_\pm^n) \ = \ & \int \dfrac{d^2r_{\per}}{(2\pi)^2}\, \dfrac{d^2s_{\per}}{(2\pi)^2} \, \dfrac{d^2t_{\per}}{(2\pi)^2} \
(u^1\pm i u^2)^n\;
\exp \left\{- \dfrac{1}{B_f} \left( t_1 \, {\vec s}_{\per}^{\;2} + \, t_2\, {\vec r}_{\per}^{\;2} + \, t_3 \,{\vec t}_{\per}^{\;2} \right) \right\}
\nonumber\\[2mm]
& \times \, \exp\left\{ i \, \dfrac{ 2 \varepsilon s_f}{B_f} \left[ ( p^1+s^1-r^1)(k^2+r^2-t^2) - (k^1+r^1-t^1)(p^2+s^2-r^2) \right] \right\} \ ,
\label{twomomenta}
\end{align}
where $n=0,1$ and $u = r,s,t\,$. Taking $\vec k_{\per} = -\vec p_{\per}$, for
$n=0$ we obtain
\begin{align}
I_{\per}(1)\ = \ \dfrac{B_f^3}{64\pi^3 T_f} \:
\exp\left[-\dfrac{ t_2 (t_1 + t_3) }{B_f\, T_f}\, \vec p_{\per}^{\:2}  \right]\ ,
\label{Iperp1}
\end{align}
while for $n=1$ we have
\begin{align}
I_{\per} (r_\pm) \ = \ & \dfrac{I_{\per}(1)}{T_f} \, (t_1+t_3)\, (p^1\pm i p^2)\ ,
\\[2mm]
I_{\per} (s_\pm) \ = \ & -\dfrac{I_{\per}(1)}{T_f} \, t_2 \, f_3^\pm\,(p^1\pm i p^2)\ ,
\\[2mm]
I_{\per} (t_\pm) \ = \ & -\dfrac{I_{\per}(1)}{T_f} \, t_2 \, f_1^\mp\,(p^1\pm i p^2)\ .
\label{iperp}
\end{align}

From the above expressions, it can be see that ${\cal
N}^{\mu\nu}_{\varepsilon f} = 0$ for $\mu\nu = 01$, 02, 10 and 20. On the
other hand, notice that for $\mu =\nu$ the expressions of ${\cal
N}^{\mu\nu}_{\varepsilon f}$ depend linearly on $\varepsilon$. Thus, the
corresponding sums over $\varepsilon = \pm 1$ also vanish. In the case of
${\cal N}^{30}_{\varepsilon f}$ and ${\cal N}^{03}_{\varepsilon f}$, one is
faced with terms that are linear in $\varepsilon$ and with integrals of
terms that are quadratic in perpendicular coordinates. For the latter, in
the integrals we can shift $r_{\per}\to r'_{\per} = r_{\per} - p_{\per}$ and
choose the 1-axis in the direction of $\vec p_{\per}$ (i.e., we can take $p^2
= 0$). Then, changing $r^{\prime\, 2}\to -r^{\prime\, 2}$, $s^2\to -s^2$ and
$t^2\to -t^2$, it is seen that the integrals change sign by changing
$\varepsilon \to -\varepsilon$, which leads to $\sum_\varepsilon {\cal
N}^{30}_{\varepsilon f} = \sum_\varepsilon {\cal N}^{03}_{\varepsilon f} =
0$. The only nonvanishing contributions to the amplitude can be obtained
from the sums
\begin{align}
\sum_{\varepsilon = \pm} \left({\cal N}^{13}_{\varepsilon f}\pm i {\cal N}^{23}_{\varepsilon f}\right)
& \ = \ - \sum_{\varepsilon = \pm} \left({\cal N}^{31}_{\varepsilon f}\pm i {\cal N}^{32}_{\varepsilon f}\right)
\nonumber \\[3mm]
& \ = \ \mp 16\,M \,I_{\per}(1)\,[1+t_1 t_3-t_2(t_1+t_3)]\, p^0 (p^1\pm ip^2) \ , \\[3mm]
\sum_{\varepsilon = \pm} {\cal N}^{12}_{\varepsilon f}  & \ = \
- \sum_{\varepsilon = \pm} {\cal N}^{21}_{\varepsilon f} \nonumber \\[3mm]
& \ = \ 16\,iM \,I_{\per}(1)\,[1+t_1 t_3-t_2(t_1+t_3)]\, p^0 p^3 \ .
\end{align}
In fact, the above results for $\sum_{\varepsilon} {\cal
N}^{\mu\nu}_{\varepsilon f}$ can be more compactly written as
\begin{align}
\sum_{\varepsilon = \pm} {\cal N}^{\mu\nu}_{\varepsilon f} & \ = \
-16\,iM I_{\per}(1)\,[1+t_1 t_3-t_2(t_1+t_3)]\,p^0\,\epsilon^{0\mu\nu\lambda}\,p_\lambda \ .
\end{align}

Turning back to the expression of $R_f^{\mu\nu}$ in Eq.~\eqref{Mmunu2}, we
are left with the integral over parallel components of $r$
\begin{align}
I_{\npar} \ = \  \int \dfrac{d^2r_{\npar}}{(2\pi)^2}\
e^{\tau_ 1 (r_{\sv[\parallel]}- p_{\sv[\parallel]})^2 +
\tau_2\, r_{\sv[\parallel]}^2 + \tau_3 (r_{\sv[\parallel]}+ k_{\sv[\parallel]})^2}\ ,
\end{align}
(notice that the result for $\sum_{\varepsilon} {\cal
N}^{\mu\nu}_{\varepsilon f}$ does not depend on $r_{\npar}$). Performing a
Wick rotation in the variable $r^0$ we get, in the pion rest frame,
\begin{align}
I_{\npar} \ = \  \dfrac{i}{4\pi \,(\tau_1 + \tau_2 + \tau_3)} \:
\exp\left[\dfrac{4\tau_1 \tau_3 (p^0)^2+ (\tau_1 + \tau_3)\tau_2\, p_{\npar}^2}{\tau_1 + \tau_2 + \tau_3}\right] \ .
\label{ipara}
\end{align}
In this way, we end up with
\begin{align}
R_f^{\mu\nu} \ = \ & 256 \, \pi^2 N_c \, \gpqq\,M\, p^0 \epsilon^{0\mu\nu\lambda}\,p_\lambda \, \dfrac{q_f^2}{B_f^2}
\nonumber\\[2mm]
& \times \, \int_0^\infty d\tau_1 \int_0^\infty d\tau_2 \int_0^\infty d\tau_3 \:
e^{-(\tau_1+\tau_2+\tau_3)M^2}I_{\per}(1)\,I_{\npar}\,[1+t_1 t_3-t_2(t_1+t_3)] \ ,
\end{align}
which, after replacing the integrals $I_{\per}(1)$ and $I_{\npar}$ by the
corresponding expressions in Eqs.~\eqref{Iperp1} and~\eqref{ipara}, leads to
the result in Eq.~\eqref{Mmunu_final}.

\section{Polarizations of outgoing photons}
\label{app.polB0} 

The polarization vectors $\epsilon^\mu(p,\pm 1)$ for an on-shell photon with
momentum $\vec p$ are given by
\begin{align}
\epsilon^0(p,\pm 1) \ = \ & 0 \ , \nonumber \\[2mm]
{\vec \epsilon}\,(p,\pm 1) \ = \ &
\dfrac{e^{\pm\, i\phi}}{\sqrt{2}\,|{\hat p}_{\per}|}\,
\big( i \hat p^2 \mp \hat p^1\hat p^3 \, ,
- i \hat p^1 \mp \hat p^2\hat p^3\, ,
\pm \,|\hat p_{\per}|^2 \big) \ ,
\label{polvecsB0}
\end{align}
where $\hat p = \vec p/|\vec p\,|$, $\hat p_{\per} = \vec p_{\per}/|\vec p\,|$ and $\phi$ is an arbitrary
phase. They satisfy
\begin{equation}
\epsilon(p,\pm 1)\cdot\epsilon(p,\pm 1)^\ast = -1\, , \quad
\epsilon(p,+1)\cdot\epsilon(p,-1)^\ast = 0\, , \quad
\epsilon(p,\pm 1)\cdot p = 0\, , \quad
\vec \epsilon\,(p,\pm 1)\cdot \vec p = 0\, ,
\end{equation}
as well as
\begin{equation}
\sum_{\lambda_p=\pm}
\epsilon^\mu(p,\lambda_p)\,\epsilon^\nu(p,\lambda_p)^\ast
\ = \ -\, \eta^{\mu\nu}\, +
\,\dfrac{p^\mu {\tilde p}^\nu +{\tilde p}^\mu p^\nu}{p\cdot \tilde p} \ ,
\end{equation}
where ${\tilde p}^\mu = (p^0,-\vec p\,)$.

\bibliography{pi0to2g_bib}

\end{document}